\documentclass[preprint,10pt,3p,twocolumn]{elsarticle}

\usepackage{graphicx} % Include figure files
\usepackage{epstopdf} %
\usepackage{dcolumn} % Align table columns on decimal point
\usepackage{bm} % bold math
\usepackage{ulem} %pretty font effects
\usepackage{threeparttable}

\usepackage{amsmath}
\usepackage{array}
\usepackage{tabulary}
\usepackage{dcolumn}% Align table columns on decimal point
\usepackage{multirow}
\usepackage{rotating}
\usepackage[usenames,dvipsnames]{color}

\makeatletter

\newcommand{\Rmnum}[1]{\expandafter\@slowromancap\romannumeral #1@}
\makeatother

\journal{Materials Letters}

\begin{document}

\begin{frontmatter}

\title{Effective Hamiltonian Methods for Predicting the Electrocaloric Behavior of BaTiO$_{3}$}

\author[isu,imr]{S. P. Beckman\corref{cor1}}
\ead{sbeckman@iastate.edu}
\author[isu]{L. F. Wan}
\author[isu]{Jordan A. Barr}
\author[isu,imr]{Takeshi Nishimatsu}

\cortext[cor1]{Corresponding author}

\address[isu]{Department of Materials Science and Engineering, Iowa State University, Ames, Iowa 50011}
\address[imr]{Institute for Materials Research (IMR), Tohoku University, Sendai 980-8577, Japan}

\begin{abstract}
The perovskite crystal BaTiO$_3$ is modeled using a first-principles based effective 
Hamiltonian and molecular dynamics simulations are performed to estimate the 
pyroelectric response.  The electrocaloric temperature change, $\Delta T$, is 
calculated for different temperatures and externally applied electric fields.  
It is found that it is possible to achieve a large $\Delta T$, around 5-6~K, 
for a relatively small electric field gradient, less than 100~kV/cm, if 
the applied fields have a small absolute magnitude.  
\end{abstract}

%\pacs{64.60.De, 77.80.B-, 77.84.-s}

\end{frontmatter}

\section{Introduction \label{introduction}}

A pyroelectric crystal develops a spontaneous electrical polarization when its temperature 
changes.~\cite{lines1977}  Its enthalpy can be expressed as $H=E-{\cal E}P$, where $E$ is the 
internal energy, ${\cal E}$ is an applied electric field, and $P$ is the polarization.  
It is possible to cycle the temperature and electric field, to drive the crystal through an 
order/disorder phase transition, such as the ferroelectric/paraelectric transition in 
perovskite crystals, to convert between heat and electric energies.  
The phenomena known as the electrocaloric effect (ECE), in which electrical energy is 
used to induce a temperature change, is directly related to this thermodynamic cycle.  
The ECE is not only interesting from a fundamental science perspective, but also holds 
great potential for future technologies such as 
solid-state refrigeration.  

The ECE was first discovered in 1930~\cite{kobeko1930}, but there were few studies 
pursued until ferroelectric crystals, with permanent dipoles, were discovered.  
The initial studies focused on bulk ceramics including 
Rochelle salt~\cite{wiseman1963}, SrTiO$_{3}$~\cite{hegenbarth1961}, 
BaTiO$_{3}$ (BTO)~\cite{lawless1980},  
and doped alkali halides~\cite{kanzig1964, shepherd1965, kuhn1965}; however, 
the engineering potential of these crystals was never realized.  

In 2006 Mischenko \textit{et al}.\ discovered the ``giant electrocaloric effect'' in Zr-rich 
PbZr$_{0.95}$Ti$_{0.05}$O$_{3}$ (PZT) thin films.~\cite{mischenko2006}  By using a thin film 
it was possible to apply ultra-high electric fields, to achieve an ECE $\Delta T$ of 12~K.  
The discovery of the giant ECE has revitalized the field and has inspired numerous recent studies 
investigating the impact of ultra-high electric fields on the pyroelectric response of 
other ferroelectric materials.    

The initial work focused on Zr-rich PZT,~\cite{mischenko2006} and although this alloy is still being 
investigated,~\cite{bellaiche2008,lwmartin} there has also been a great deal of interest in other 
perovskite crystals including 
$\left(1-x\right)$PbMg$_{1/3}$Nb$_{2/3}$O$_{3}$--$x$PbTiO$_{3}$~\cite{zhang2006,hagberg2008},
BTO~\cite{cao2009,akcay2007,bai2010,2bai2010,kar-narayan2010,fang2011},
BaTiO$_{3}$--SrTiO$_{3}$ superlattices~\cite{juqiu2009},
and SrBi$_2$Ta$_2$O$_9$ layered oxides~\cite{chenhao2009}.  
In addition there has been much interest in the development of 
polymer based ferroelectrics, such as 
poly(vinylidene fluoride-trifluoroethylene) 
and poly(vinylidene fluoride-trifluoroethylene-chlorofluoroethylene).~\cite{neese2008,sglu2009}

In this paper we will focus on the BTO compound for a variety of reasons: 
it is a well studied archetypical perovskite crystal that is relatively easy to produce, 
it exhibits an excellent pyroelectric response,
it does not contain the toxin lead, its ferroelectric properties have the potential to be tailored by 
alloying or the creation of superlattice structures.  
The ECE in BTO has been previously studied as a 
thin film\cite{cao2009,akcay2007} and as a thick film in a multi-layer stack.~\cite{bai2010,2bai2010}  
Surprisingly, off-the-shelf BTO multilayer capacitors have been shown to exhibit a reasonable 
electrocaloric response.~\cite{kar-narayan2010}    
There also has been a study of the properties of BTO nanoparticles with core-shell 
geometries.~\cite{fang2011}  

Previous theoretical studies of BTO have largely relied on 
thermodynamic models, for example the Ginzburg-Landau-Devonshire 
model.~\cite{akcay2007, bai2010,2bai2010, fang2011,juqiu2009,lwmartin}  
While these have proven successful, they require a large amount of data to accurately parameterize the model.  
One alternative approach was demonstrated by Cao and Li in which they created a transverse Ising model (TIM) 
to study the ECE in BTO thin-films.  
They predict that compressive epitaxial strains result in increasing the temperature where the 
ECE is maximum and tensile strains reduce the temperature.~\cite{cao2009} The TIM model is particularly 
interesting because in addition to the spin exchange interactions, taken up to four spins, the effect of quantum 
fluctuations were also included.  It is reported that increasing the strength of the quantum fluctuations results in 
shifting the ECE to higher temperatures.  
Bellaiche \textit{et al.}~have used a molecular dynamics approach 
to investigate nanodots of
$\mbox{Pb}\left(\mbox{Zr}_{0.4}\mbox{Ti}_{0.6}\right)\mbox{O}_{3}$.  
They report the time dependency of the temperature when an alternating
electric field is applied.~\cite{bellaiche2008}  
It is discovered that it may be possible to tune the ECE in nanodots by adjusting the 
depolarization field, which in practical terms would be implemented by
modifying the surface of the nanodot. 

Following this introduction the theoretical methods are presented. The
results are presented and discussed in section \ref{results}. The
manuscript concludes with a summary in section \ref{summary}.

\section{Methods \label{methods}}

Here we predict the ECE for BTO using the model Hamiltonian, 
\begin{multline}
  \label{eq:Effective:Hamiltonian}
  H^{\rm eff}%(\{\bm{u}\},\{\bm{w}\}, \eta_1,\cdots\!,\eta_6)
  = \frac{M^*_{\rm dipole}}{2} \sum_{\bm{R},\alpha}\dot{u}_\alpha^2(\bm{R})
  + \frac{M^*_{\rm acoustic}}{2}\sum_{\bm{R},\alpha}\dot{w}_\alpha^2(\bm{R})\\
  + V^{\rm self}(\{\bm{u}\})+V^{\rm dpl}(\{\bm{u}\})+V^{\rm short}(\{\bm{u}\})\\
  + V^{\rm elas,\,homo}(\eta_1,\cdots\!,\eta_6)+V^{\rm elas,\,inho}(\{\bm{w}\})\\
  + V^{\rm coup,\,homo}(\{\bm{u}\}, \eta_1,\cdots\!,\eta_6)+V^{\rm coup,\,inho}(\{\bm{u}\}, \{\bm{w}\})\\
  -Z^*\sum_{\bm{R}}\bm{\mathcal{E}}\!\cdot\!\bm{u}(\bm{R}), 
\end{multline}
which is derived in Refs.~\cite{King-Smith:V:1994,Zhong:V:R:PRB:v52:p6301:1995}.  
The true atomic structure has properties that are determined by the 
complex chemical bonding between the atoms, but in the model system the complexity is 
reduced; the collective atomic motion is 
coarse-grained by local softmode
vectors $\bm{u}(\bm{R})$ and
local acoustic displacement vectors $\bm{w}(\bm{R})$
of each unit cell at $\bm{R}$ in a simulation supercell as well as the 
homogeneous strain components, $(\eta_1,\eta_2,\eta_3, \eta_4,\eta_5,\eta_6)$.
The terms in the Hamiltonian bear a physical significance:  
 $\frac{M^*_{\rm dipole}}{2} \sum_{\bm{R},\alpha}\dot{u}_\alpha^2(\bm{R})$ and 
  $\frac{M^*_{\rm acoustic}}{2}\sum_{\bm{R},\alpha}\dot{w}_\alpha^2(\bm{R})$ is the 
  kinetic energy possessed by the local soft modes and 
  the local acoustic displacement vectors, 
  $V^{\rm self}(\{\bm{u}\})$ is the local mode self energy, 
  $V^{\rm dpl}(\{\bm{u}\})$ is the long-ranged dipole-dipole interaction, 
  $V^{\rm short}(\{\bm{u}\})$ is the short-ranged interaction between local soft modes,
  $V^{\rm elas,\,homo}(\eta_1,\cdots\!,\eta_6)$ is the elastic energy from homogeneous strains, 
   $V^{\rm elas,\,inho}(\{\bm{w}\})$ is the elastic energy from inhomogenous strains, 
   $V^{\rm coup,\,homo}(\{\bm{u}\}, \eta_1,\cdots\!,\eta_6)$ is the coupling between the local soft modes and the homogeneous strain, 
$V^{\rm coup,\,inho}(\{\bm{u}\}, \{\bm{w}\})$ is the coupling between the soft modes and the inhomogeneous strains, and 
  $-Z^*\sum_{\bm{R}}\bm{\mathcal{E}}\!\cdot\!\bm{u}(\bm{R})$ is the electric enthalpy.  
Each of the terms in Eq.\ref{eq:Effective:Hamiltonian} is expanded to allow the Hamiltonian 
to be expressed using the parameterization given in 
Refs.~\cite{King-Smith:V:1994,Zhong:V:R:PRB:v52:p6301:1995,Nishimatsu:feram:PRB2008,Waghmare:C:B:2003,Nishimatsu.PhysRevB.82.134106}.  
The same set of parameters in Ref.~\cite{Nishimatsu.PhysRevB.82.134106} for BaTiO$_3$ is used in this letter.
Values of the 25 parameter are also given as Supplementary Data to this letter.
The details of the molecular dynamics method are explained in 
Refs.~\cite{Nishimatsu:feram:PRB2008,Waghmare:C:B:2003}.  
The method described here is encoded in the software package \texttt{feram}, which 
is distributed freely to the scientific community.\footnote{The MD code ``\texttt{feram}'' can be downloaded 
from \texttt{http://loto.sourceforge.net/feram/} as free software.}  

%\begin{table}
%  \caption{The effective Hamiltonian parameters for BaTiO$_{3}$ used in this model 
%from Ref.~\cite{Nishimatsu.PhysRevB.82.134106}}
%  \label{tab:parameters}
%  \centering
%  \begin{tabular}{llll}
%    \hline
%    \hline
%
%    Parameter & Value & Parameter & Value \\
%    \hline
%    $p$       [GPa]              & $-0.005T$   & $\epsilon_\infty$           &   6.87  \\
%    $a_0$    [\AA]              &  3.986       & $\kappa_2$ [eV/\AA$^2$]    &   8.534  \\
%    $B_{11}$   [eV]            & 126.73      &  $j_1$     [eV/\AA$^2$]     &         $-2.084$  \\
%    $B_{12}$   [eV]            &  41.76      & $j_2$     [eV/\AA$^2$]     &         $-1.129$  \\
%    $B_{44}$   [eV]            &  49.24     &  $j_3$     [eV/\AA$^2$]       &           0.689 \\
%    $B_{1xx}$  [eV/\AA$^2$]      &  $-185.35$ &  $j_4$     [eV/\AA$^2$]     &         $-0.611$ \\
%    $B_{1yy}$  [eV/\AA$^2$]       &  $-3.2809$  &  $j_5$     [eV/\AA$^2$]     &           0.000 \\
%    $B_{4yz}$  [eV/\AA$^2$]     &  $-14.550$   & $j_6$     [eV/\AA$^2$]     &           0.277 \\
%    $\alpha$  [eV/\AA$^4$]      &    78.99   &  $j_7$     [eV/\AA$^2$]       &           0.000 \\
%    $\gamma$  [eV/\AA$^4$]       & $-115.48$  &  & \\
%    $k_1$     [eV/\AA$^6$]         & $-267.98$  &  & \\
%    $k_2$     [eV/\AA$^6$]         &   197.50       &  & \\
%    $k_3$     [eV/\AA$^6$]    &   830.20         &  & \\
%    $k_4$     [eV/\AA$^8$]       &   641.97     &  & \\
%    $m^*$     [amu]              &   38.24       &  & \\
%    $Z^*$     [e]              &  10.33   &  & \\
%    \hline
%    \hline
%  \end{tabular}
%\end{table}

The simulation uses a supercell of size $N = L_x\times L_y\times L_z = 16 \times 16 \times 16$
and the temperature is scanned from 250~K to 900~K at an increment of $+1$~K/step.
For each temperature increment, the system is thermalized for 20,000 time steps,
after which the properties are averaged for 80,000 time steps, using a time step of $\Delta t=2$~fs.
The simulations begin at low temperature with a $z$-polarized initial configuration that is random.
As described in Ref.~\cite{Nishimatsu.PhysRevB.82.134106},
thermal expansion is simulated by including a temperature-dependent effective 
negative pressure $p=-0.005T$~GPa. 
It has been shown in Ref.~\cite{Nishimatsu.PhysRevB.82.134106} that 
the parameterization used here offers an 
improved estimation of ferroelectric (tetragonal)
to paraelectric (cubic) transition temperature and $c/a$ ratio
as compared to earlier models~\cite{King-Smith:V:1994,Zhong:V:R:PRB:v52:p6301:1995}.
Therefore, we believe that this set of parameters also gives a good 
estimation of the pyroelectric properties.

The complex interactions, such as the interaction of domain walls 
with crystal defects and interfaces that are present in experimental systems are not 
present here. 
This makes direct comparison between computational and experimental results 
challenging; however, these calculations are intended to isolate the essential 
physics of the crystal and the pyroelectric response due to the bonds between 
the metal cations and the oxygen. It is anticipated that the fundamental 
behavior observed in these calculations will also be present in the experimental 
systems.  In principle the predicted ECE response will be observed in high purity 
single crystals of BTO, although such experimental systems are challenging to 
prepare. 

\section{Results and Discussion \label{results}}

\begin{figure}[p!]
  \centering
  % From appb:/home/t-nissie/feram/feram-e.07/BaTiO3-020-WuC-0.005-100keVcm/
  \includegraphics[width=80mm]{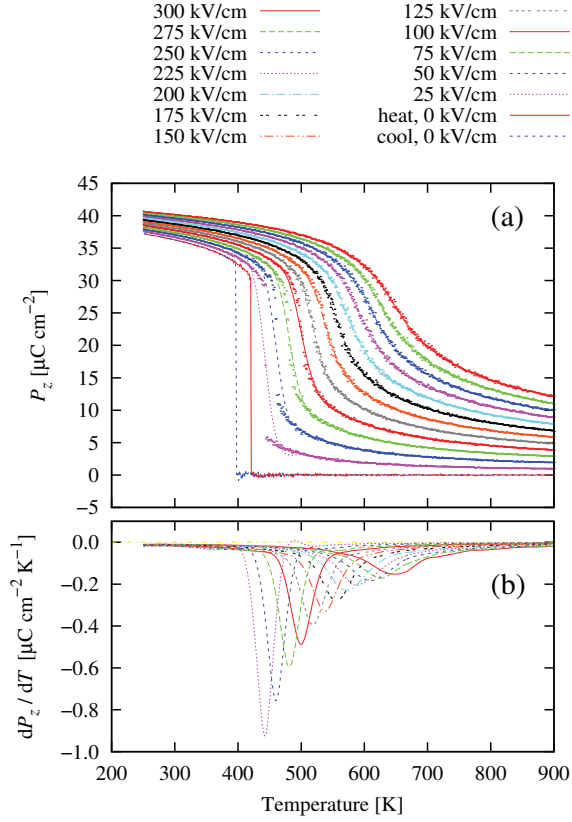}
  \caption{(color online) The calculated pyroelectric response. 
  (a) The simulated temperature dependence of the polarization along the $z$-direction
    at constant external electric field is determined.  The simulated data 
    is plotted as dots and are fitted using weighted splines, shown as lines. 
    ${\cal E}_{z}=0$ kV/cm heating-up and cooling-down results are also plotted for 
    reference. 
    (b) The pyroelectric coefficient, $\partial P_z/ \partial T$, is determined from 
    the fitted data.}
  \label{fig:T}
\end{figure}

\begin{figure}[p!]
  \centering
  % From appb:/home/t-nissie/feram/feram-e.07/BaTiO3-020-WuC-0.005-100keVcm/
  \includegraphics[width=80mm]{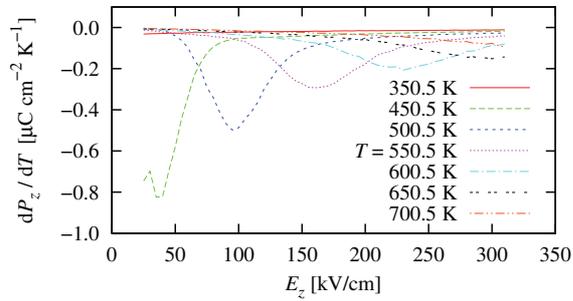}
  \caption{
    (color online)
    The pyroelectric coefficient, $\partial P_z/ \partial T$, is determined as a function of electric field
    at constant temperature from the fitted data in Fig.~\ref{fig:T}.
  }
  \label{fig:E}
\end{figure}

Simulations are performed for constant electric fields ranging from 
25 to 310~kV/cm at an increment of 5~kV/cm. 
This temperature range focuses on the ferroelectric (tetragonal) to 
paraelectric (cubic) transitions, where the largest pyroelectric response should 
be observed. 
The $z$-component of polarization, $P_z$, is calculated from the MD simulation results 
and is plotted versus $T$ at constant ${\cal E}_z$ in Fig.~\ref{fig:T}(a).
The data are fitted with weighted splines to allow for the determination of 
$\partial P_z/ \partial T$, which is shown in Fig.~\ref{fig:T}(b) and Fig.~\ref{fig:E}.
Cooling-down simulations are also performed 
and hysteresis in temperature is found when ${{\cal E}_z} \leq 25$ kV/cm,
as seen in the ${{\cal E}_z}=0$ kV/cm simulations in Fig.1(a).
In the simulations, the phase-transition order changes from 
first to second around ${{\cal E}_z}=25$ kV/cm. 
To focus our theoretical effort on the determination of the pyroelectric response without 
including the additional complexity of the latent heat of transformation, 
only fields ${{\cal E}_z}\ge60$ kV/cm are used in the present calculations.

\begin{figure}[p!]
  \centering
  % From appb:/home/t-nissie/feram/feram-e.07/BaTiO3-020-WuC-0.005-100keVcm/
  \includegraphics[width=80mm]{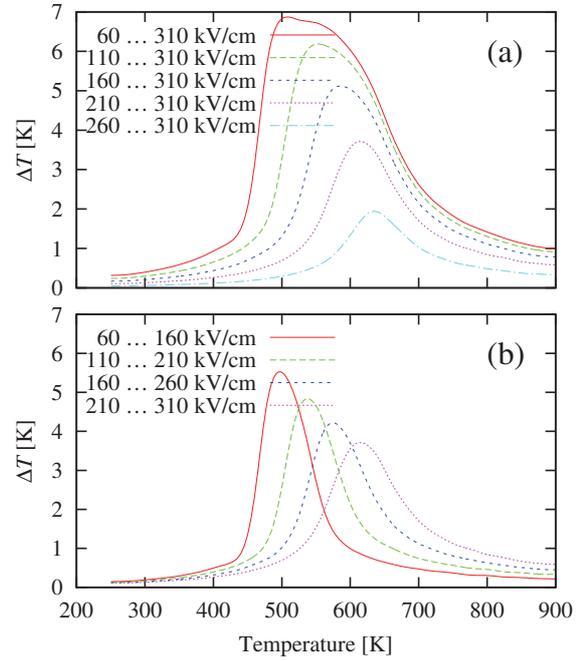}
  \caption{  
    (color online)
    The predicted electrocaloric effect estimated from Eqn.~\ref{eq:DeltaT}.  
    The integral range ${{\cal E}_1}$~\ldots~${{\cal E}_2}$ is specified in the 
    plot legend. 
    (a) The value of ${{\cal E}_2}$ is held fixed at 310~kV/cm and 
    ${{\cal E}_1}$ is varied to give the $\Delta {{\cal E}}$ the values 
    250, 200, 150, 100, and 50~kV/cm.  
    (b) The values of ${{\cal E}_1}$ and ${{\cal E}_2}$ are changed together 
    to maintain a fixed value of $\Delta {{\cal E}}=100$~kV/cm.}
  \label{fig:DeltaT}
\end{figure}

Following Ref.~\cite{mischenko2006},
the electrocaloric refrigeration $\Delta T$ is estimated using 
\begin{equation}
  \label{eq:DeltaT}
  \Delta T = \frac{-1}{C_v} \int^{{\cal E}_2}_{{\cal E}_1}T\frac{\partial P_z}{\partial T} d{{\cal E}}_z
\end{equation}
as shown in Fig.~\ref{fig:DeltaT}. For heat capacity, experimental,
temperature and electric-field independent value of
$C_v=5.84~\text{g}\cdot\text{cm}^{-3}
 \times
 0.434~\text{J}\cdot\text{g}^{-1}\cdot\text{K}^{-1} $ or  
$2.53~\text{J}\cdot\text{cm}^{-3}\cdot\text{K}^{-1}$
is used in this estimation~\cite{Yi2004135}; 
the coarse-graining used in the effective Hamiltonian 
method makes it difficult to correctly estimate the heat capacity.   
The integral in Eqn.~\ref{eq:DeltaT} ranges over 
$\Delta {{\cal E}}={{\cal E}_2}-{{\cal E}_1}$.  The 
fields used for this integral are specified in the legends 
of Fig.~\ref{fig:DeltaT} as, ``${\cal E}_1$~\ldots~${{\cal E}_2}$.'' 

Because the phase transition becomes increasingly diffuse as the electric field is 
increased the greatest pyroelectric response, 
$\partial P_z/ \partial T$ shown in Fig.~\ref{fig:T} (b), 
occurs at lower electric fields.  
In Fig.~\ref{fig:DeltaT} frame (a) the field gradient $\Delta {\cal E}$ is varied and 
${\cal E}_2$ is held constant at 310~kV/cm. The magnitude of the ECE $\Delta T$ is 
proportional to the magnitude of the field gradient, $\Delta {\cal E}$.  
This is entirely consistent with the experimental observations. 
The temperature where the maximum $\Delta T$ occurs 
is determined by the isothermal curve in Fig.~\ref{fig:T} that 
makes the greatest contribution within the field range bracketed 
by ${\cal E}_1$ and ${\cal E}_2$.  
For example, over the range ${\cal E}_1=160$ and ${\cal E}_2=310$~kV/cm 
the $T=550.5$ and 600.5~K curves have the greatest magnitude in Fig.~\ref{fig:E} 
and the $\Delta T$ versus $T$ curve, in Fig.~\ref{fig:DeltaT} (a), peaks just 
above $T=550$~K.

It is observed by Akcay \textit{et al.}~that not only is the field 
gradient, $\Delta {\cal E}$, important but also the absolute magnitude of the 
fields ${\cal E}_1$ and ${\cal E}_2$.  
In Fig.~\ref{fig:DeltaT} (b) the field gradient is held constant at 
$\Delta {\cal E}=100$ and ${\cal E}_1$ is varied from 
60 to 210~kV/cm.  At larger absolute fields the diffuse nature of 
the phase transition results in a decrease in the magnitude of the 
ECE $\Delta T$ as well as an increase in the temperature where 
$\Delta T$ is maximum.  As a matter of practice this means that 
the ECE will always occur above the Curie temperature.  
Comparing the results in Fig.~\ref{fig:DeltaT} (a) and (b) it is 
observed that by keeping ${\cal E}_1$ small it is possible 
to achieve large $\Delta T$ for small $\Delta {\cal E}$, 
and when ${\cal E}_1$ becomes large it is necessary to 
have a large $\Delta {\cal E}$ to have comparable $\Delta T$.  
The implication of this observation is that compounds with 
antiferroelectric ordering, which require large 
${\cal E}_1$,\cite{mischenko2006} may not be the optimal choice 
for producing a large ECE.  

\section{Summary \label{summary}}

In summary, we have demonstrated the use of our MD method \texttt{feram}
to estimate the ECE as functions of temperature and external fields.
The freely available method
used here can be parameterized using 
first-principles results alone, and therefore can be extended to investigate systems 
that lack sufficient experimental data to produce an accurate thermodynamical model.  
We have demonstrated the effect of varying electric fields on the archetypical 
pyroelectric, BTO, and have found that to produce a large ECE for a 
relatively small field gradient, it is preferred to not have an 
antiferroelectric ground state, which will allow the use of modest fields.  
While admittedly the external electric fields consider here may be 
slightly high compared to the breakdown voltage of polycrystalline 
BaTiO$_{3}$; advancements in synthesis techniques, such as 
vapor deposition processes, hold the promise of creating affordable 
high quality ferroelectric thin films that can withstand these 
electric fields.  

\section{Acknowledgements}

The work of SPB, JAB, and LFW was supported by the National Science
Foundation (NSF) through Grant DMR-1037898 and DMR-1105641.
SPB would like to thank
the International Collaboration Center at the Tohoku University
Institute for Materials Research for their support summer 2011. The
work of TN was supported by Japan Society for the Promotion of Science
(JSPS) through KAKENHI 23740230. Computational resources were provided
by the Center for Computational Materials Science, Institute for
Materials Research (CCMS-IMR), Tohoku University. 

%\bibliographystyle{model1-num-names}
%\bibliography{SPB-pyrobib,ferroelectrics}

\end{document}